\documentclass[twocolumn,aps,prl,amsmath,a4paper,superscriptaddress]{revtex4}

\usepackage{graphicx}    

\graphicspath{{Figures/},{./}}

\begin{document}


\title{Synchronization Model for Stock Market Asymmetry}

\author{Raul Donangelo}
\email{donangel@if.ufrj.br} 
\affiliation{Instituto de Fisica da UFRJ, Caixa Postal 68528,
  21941-972 Rio de Janeiro, Brazil}

\author{Mogens H. Jensen}
\email{mhjensen@nbi.dk}
\affiliation{The Niels Bohr Institute, Blegdamsvej 17, DK-2100
  Copenhagen, Denmark}

\author{Ingve Simonsen}
\email{Ingve.Simonsen@phys.ntnu.no}
\affiliation{Department of physics, Norwegian University of Science
  and Technology (NTNU), NO-7491 Trondheim, Norway}
\affiliation{NORDITA, Blegdamsvej 17, DK-2100 Copenhagen {\O}, Denmark}

\author{Kim Sneppen}
\email{sneppen@nbi.dk}
\affiliation{The Niels Bohr Institute, Blegdamsvej 17, DK-2100
  Copenhagen, Denmark}

\date{\today}

\begin{abstract}
  The waiting time needed for a stock market index to undergo a given
  percentage change in its value is found to have an up-down
  asymmetry, which, surprisingly, is not observed for the individual
  stocks composing that index.  To explain this, we introduce a market
  model consisting of randomly fluctuating stocks that occasionally
  synchronize their short term draw-downs.  These synchronous events
  are parameterized by a ``fear factor'', that reflects the occurrence
  of dramatic external events which affect the financial market.
\end{abstract}

\maketitle

\begin{figure}
\includegraphics*[width=7cm]{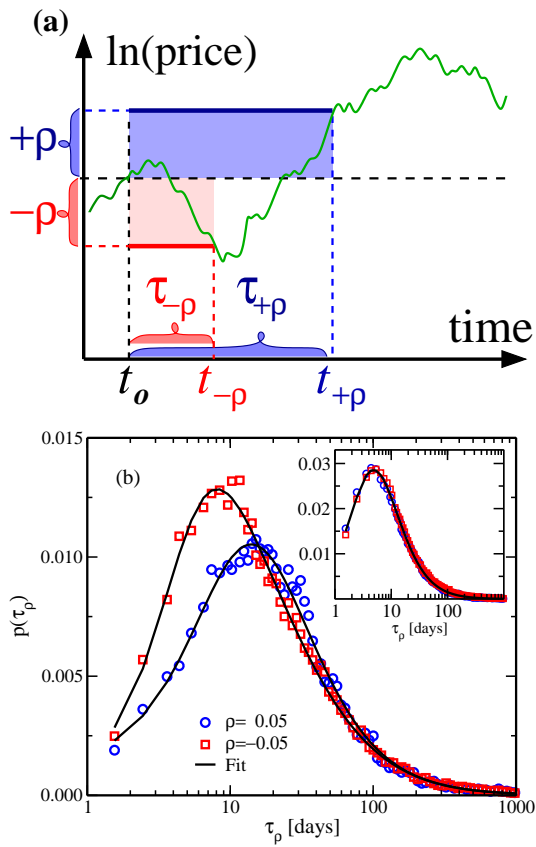}\\*[2mm]
\includegraphics*[width=7cm]{fig1b.eps}
\caption{(Color online) (a): Schematic picture of typical stock or
  index log-price variations with time.  The predetermined return
  levels for gain (blue area)/loss (red area) are set to
  $+\rho>0$/$-\rho<0$.  The corresponding investment waiting times
  ($\tau_\rho$) are found by estimating where the horizontal $\pm\rho$
  lines cross the logarithmic price curve for the {\it first} time
  ($t_\rho$), resulting in $\tau_\rho = t_\rho-t_0$.  (b): The 
  panel shows histograms of the inverse statistics for the DJIA
  obtained on the basis of the empirical daily close data covering its
  entire history of 110 years.  The red data points are obtained using
  a loss level of $\rho = - 0.05$ and the blue points are obtained
  using a gain level of $\rho = + 0.05$ and both distributions are
  normalized.  Note the clear asymmetry between the loss and the gain
  statistics.  The full curves are fits using generalized inverse
  Gamma distributions \cite{optihori,gainloss}.  The inset is the
  distribution obtained from using the same procedure on the
  individual stocks of the DJIA, and subsequently averaging over the
  stocks.  Notice that the asymmetry is absent for individual stocks.}
\label{fig1}
\end{figure}

The value of stocks varies from day to day, both relative to each
other but also due to collective movements of the overall market.
These variations of the market presumably reflect the psychological
state of the surrounding society as affected by current events.  An
analysis technique based on inverse statistics has recently been
applied to study the variation of stock indices, single stocks and
exchange rates~\cite{optihori,gainloss,invfx}.  In the time-dependent
inverse statistics approach, one fixes a predetermined level of return
($\rho$), and, as explained in Fig.~\ref{fig1}(a), asks for the
waiting time needed to reach this level for the {\em first} time.
Averaging over many investment events, results in a histogram of
waiting times.

The DJIA is an average over $30$ of the most important stocks of the
US-market chosen from different sectors of the industry.  When the
gain($+$)/loss($-$) return levels are set to $\rho=\pm 5\%$,
Fig.~\ref{fig1}(b) shows the histograms obtained for the DJIA daily
closing values over its entire 110 years history.  The histograms
possess well defined and pronounced maxima, the so-called {\em optimal
  investment horizons}~\cite{optihori}, followed by long power law
tails.  These $1/t^{\alpha}$-tails, with $\alpha\approx3/2$, are well
understood, and are a consequence of the uncorrelated increments of
the underlying asset price
process~\cite{Book:Bouchaud-2000,Book:Mantegna-2000,Book:Hull-2000,Book:Johnson-2003}.
However, the interesting observation to be made from
Fig.~\ref{fig1}(b) is that the optimal investment horizons of the same
magnitude, but opposite signs, are {\em different}.  More
specifically, for $\rho=5\%$ the maximum occurs after around
$15$~days, while for the mirrored (loss) case of $\rho=-5\%$ it occurs
at about $8$~days.  Thus the market as a whole, as monitored by the
DJIA, exhibits a fundamental {\em gain-loss asymmetry}.  Other
indices, such as SP500 and NASDAQ, also show this
asymmetry~\cite{Johansen}, while, for instance, foreign exchange data
do not~\cite{invfx}.

The striking paradox is that a similar well-pronounced asymmetry is not found 
for any of the individual stocks that compose the DJIA~\cite{Johansen}.  
This can be observed from the insert of Fig.~\ref{fig1}(b), which shows the 
results of applying the same procedure, individually, to these stocks, and 
subsequently averaging over them to improve the statistics.  
The figure illustrates that single stocks show inverse statistic histograms 
that are similar to the DJIA index, but with the important difference that 
there is no asymmetry between gains and losses.  
How is it possible that the index exhibits a pronounced asymmetry while the 
individual stocks do not?

Motivated by numerous empirical studies, the classic assumption in
theoretical finance is that stock and market prices are approximated
by a {\em geometrical Brownian
  motion}~\cite{Book:Bouchaud-2000,Book:Mantegna-2000,farmer}, {\it
  i.e.}  the logarithm of the stock price is consistent with a
standard, unbiased, random walk.  We will adopt this assumption in the
following.  Moreover, it will be assumed, as is consistent with
empirical findings \cite{Book:Bouchaud-2000,Book:Mantegna-2000}, that
the stock-price increments are small compared to the stock price level
ensuring that the logarithmic return distribution is symmetric.  Under
these assumptions a single stock will not show any gain-loss
asymmetry.  Thus, we will have to understand how the average of many
individual (log-normal distributed) stocks can collectively add up to
exhibit an gain-loss asymmetry for the resulting index that was not
present for the constituting single stocks.  The prime idea is to
introduce occasional {\em synchronous events} among the individual
stocks, {\it i.e.} a collective phenomenon.  To this end, we introduce
a model consisting of $N$ log-normally distributed stocks of price
$S_i(t)$ ($i=1,\ldots,N$) that, at each (discrete) time step, $t$, can
adjust their logarithmic prices $s_i(t)=\ln S_i(t)$ up or down by a
certain amount $\delta>0$, which for simplicity is assumed to be
constant, but with the direction $\varepsilon_i(t)=\pm 1$ chosen
randomly:
\begin{align}
  s_i(t+1) = s_i(t) + \varepsilon_i(t) \delta\;.
\label{logprice}
\end{align}
Notice that this update rule implies that the logarithmic return is
\begin{align}
r_i(t) = \ln(S_i(t+1)/S_i(t))=\varepsilon_i(t) \delta\;.
\label{logreturn}
\end{align}  
With such price process applying to all the component stocks, the
value of the (price-weighted) stock index (like the DJIA) is
calculated according to:
\begin{align}
 I(t) = \frac{1}{d(t)} \sum_{i=1}^N S_i(t) \;=\;
          \frac{1}{d(t)} \sum_{i=1}^N \exp s_i(t)\;,
\label{index1}
\end{align}
where $d(t)$ denotes the divisor of the index (at time $t$). This
quantity is adjusted over time to take into account structural changes
on the index, like for instance stock spits and mergers. However, we
have in this work, for simplicity, not considered such possibilities
and instead fixed its value to $d(t)=N$ (the initial value originally
used by the DJIA). This price-weighted way of calculating the index,
as already mentioned, is consistent with the DJIA, but is, however,
more the exception than the rule.  A more common scenario is to
construct the index from the sum of the capitalizations of the
constituent companies (a market capitalization weighted index).  This
is obtained by summing, for each company, the product of the number of
shares and the share price.  This is the way that {\it e.g.} the
NASDAQ and the SP500 indices are calculated.  Notice, however, that
the gain-loss asymmetry does not depend on the way that the index is
calculated, since the same type of behavior is found in both of them.

With these definitions in place, the increments of the index itself
between two consecutive days can be written as
\begin{align}
 \Delta I(t) = & I(t+1)-I(t)
      = \frac{1}{d(t)}
        \sum_{i=1}^N S_i(t) \left[ e^{\varepsilon_i(t)\delta}-1\right],
\label{index2} 
\end{align}
where one simply has substituted the expression (\ref{logprice}) into
the definition (\ref{index1}). Notice that this expression contains a
price dependent weight factor that comes about due to the geometric
Brownian motion assumption for the stock prices.  In the limit of
small $\delta$, the expression for the increments of the index may be
well approximated by its first order expansion,  so that
\begin{align}
  \Delta I(t) 
  \simeq \frac{\delta}{d(t)}  \sum_{i=1}^{N}   
  \varepsilon_i(t) \, S_i(t), 
  \qquad  \delta\ll 1. 
\end{align}

\begin{figure}
  \includegraphics*[width=7cm]{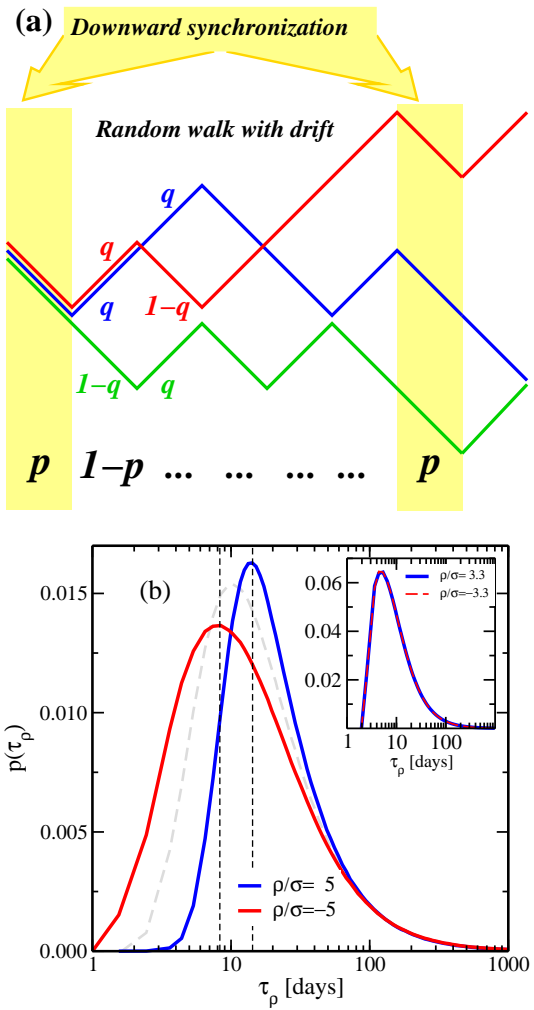}\\*[4mm]
  \includegraphics*[width=7cm]{fig2b.eps}
  \caption{(Color online) The asymmetric synchronous model.  (a): The
    panel illustrates the time evolution of three stocks, which
    fall simultaneously with probability $p$, or move as (biased)
    random walkers with probability $1-p$ (see the text for additional
    details).  (b): The inverse statistics
    obtained within this model, for an index consisting of $N=30$
    stocks and a ratio $\rho/\sigma=5$, where $\rho$ is the return
    level and $\sigma$ denotes the daily volatility of the index.  In
    the model the index volatility, $\sigma$, should reflect the
    observed $1$\% daily volatility of the DJIA, and the
    $\rho/\sigma=5$ therefore corresponds to $\rho=5$\% in
    Fig.~\protect\ref{fig1}.  A fear factor of $p=0.05$ was chosen to
    reproduce the positions of the two asymmetric maxima appearing in
    Fig.~\ref{fig1}(b) and indicated by dashed vertical lines.  The
    dashed thick line is the result for a fear-factor parameter $p=0$,
    in which case the asymmetry vanish.  As in
    Fig.~\protect\ref{fig1}(b), the inset shows the loss and gain
    distributions for the individual stocks in the model.  Notice,
    that here the asymmetry is also absent.}
\label{fig2}
\end{figure}

Synchronization is introduced into the model via simultaneous down
movements of all stocks at some time steps.  The frequency of such
events is given by a ``fear-factor'' parameter $p$.  Therefore, at
each time step, with probability $p$ {\em all} stocks move down
synchronously, {\it i.e.} for that time step $\varepsilon_i=-1$ for
all $i$, and with probability $1-p$ each stock makes an independent
and random adjustment to its logarithmic stock price.  The process is
illustrated in Fig.~\ref{fig2}(a).  To guarantee that the logarithmic
prices of the individual stocks behave like standard random walks
without any drift, the forced down movements is compensated with a
slight tendency of up-movements in the calm periods between
synchronized downwards events.  That is, on a day without synchronized
movements, the chance for a stock to move up $q$ is slightly bigger
than its chance to move down, $1-q$.  These latter situations
correspond to $\varepsilon_i=1$ and $\varepsilon_i=-1$, respectively.
Notice that the precise value of $q$ depends on the fear-factor $p$,
and is determined by equating the probabilities of up and down
movements.  The probability to move up, $(1-p) \cdot q$, must
therefore equal the probability to move down, $p+(1-p)\cdot (1-q)$,
implying that
\begin{align}  
  q = \frac{1}{2(1-p)}.
\end{align}  
We stress that a single stock generated this way, by construction,
will show no gain-loss asymmetry.  This is exemplified by {\it e.g.}
the inset to Fig.~\ref{fig2}(b).

The fear factor parameter $p$ reflects a collective anxiety state of
the investors, likely triggered by unexpected events.  But how often
do such events occur?  We have found that a value $p=0.05$, that
corresponds to one collective event every month or so, reproduces the
empirical asymmetry.  Fig.~\ref{fig2}(b) depicts the inverse
statistics of the model shown in Fig.~\ref{fig2}(a).  In obtaining
these results we used a fear factor of $p=0.05$, $N=30$ stocks, and
the return level was set to $\rho/\sigma=\pm5$, where $\sigma$ denotes
the daily volatility of the index~\footnote{The input parameter to the
  model was in practice the single stock volatility (and not that of
  the index). The volatility of the statistically identical single
  stocks was adjusted, given the other parameters of the model, to
  produce the desired volatility of the index $\sigma$.}.  For the
DJIA, the daily volatility is about $1\%$, and hence the model value
$\rho/\sigma=5$ should be comparable to the $\rho=5\%$ for the
DJIA-index used in obtaining the empirical results of
Fig.~\ref{fig1}(b). Moreover, it should be noted that when $p=0$ no
asymmetry is predicted by the model (dashed line in
Fig.~\ref{fig2}(b)).

Fig.~\ref{fig2}(b) shows that the model results in a clear gain-loss
asymmetry that is qualitatively very similar to what is found
empirically for the DJIA (cf. Fig.~\ref{fig1}(b)).  In particular, the
empirical peak positions are determined rather accurately by the
model, as indicated by the vertical dashed lines in
Fig.~\ref{fig2}(b).  A detailed comparison of the shapes of the
empirical and the modeled inverse statistics curves reveals some minor
differences, especially for short waiting times.  One could find
simple explanations for these differences, such as the fact that the
model does not consider a realistic jump size distribution, or even
that it does not include an ``optimism factor'' synchronizing
draw-ups.  This would result in a wider $\rho>0$ distribution for
short waiting times, and additionally would lower the value of the
maximum.  However, we have chosen not to include any of these
additional issues into the phenomenological model, in order to keep it
as simple and transparent as possible, and since it serves well to our
main aim, which is to address the origin of the asymmetry.

Our model opens for additional investigations into the effects of the
many small synchronous events in the market (``mini-crashes'').  In
particular, we have studied the probability that the DJIA index goes
down, respectively up, over a day and have found that there is $9\%$
larger probability to go down than to go up.

This is in perfect agreement with the model where the index has a
larger probability to go down because of the synchronizing draw-down
events, as quantified by the fear-factor $p$.  Moreover, we have found
overall quantitative agreement between the empirical DJIA data and the
model (with the parameters given above) for the probability of moving
up/down $M$ consecutive days.  The peak positions in
Figs.~\ref{fig1}(b) and ~\ref{fig2}(b) are obviously related to the
value chosen for $\rho$.  As $\rho$ increases, the peaks move to
longer times~\cite{optihori}, and their amplitudes decrease.

One might speculate whether the observed asymmetry could be used to
generate profit.  It cannot (we believe)!  A call (put) option
contract gives the holder the right to buy (sell) and obliges the
writer to sell (buy) a specified number of shares at a specified
strike price, any time before its expiry date.  If we implemented a
strategy based on a put option at current price eight days from now
(corresponding to the maximum loss curve), and a call option at
current price 15 days from now (corresponding to maximum probability
of gain curve), one can demonstrate that the expected long term gain
is mathematically identical to a straight forward hold position.
Obviously, the cost of buying the options and any additional
transaction costs, would render the use of our observed asymmetry
unprofitable.

The asymmetry of markets reflects an inherent difference between the
value of money and the value of stocks, where crashes reflect the
tendency of people to believe in money, rather than stocks, during
crises.  In this perspective it is interesting to notice that it is
the index, {\it i.e.} the value of all stocks, that is systematically
vulnerable relative to the more fluid money.  The buying power of
money is complementary to value of stocks \cite{menger}, and thus
exhibits a mirrored asymmetry with a tendency of an increased buying
power for money relative to index at short times.  In periods of fear,
people prefer money as the more certain asset, while calm periods are
characterized by random reshuffling of agent's stock assets with a
tendency to push stock values upwards.

We conclude that the asymmetric synchronous market model captures
basic characteristic properties of the day-to-day variations in stock
markets.  The agreement between the empirically observed data here
exemplified by the DJIA index and the parallel results obtained for
the model gives credibility to the point that the presence of a
``fear-factor'' is a fundamental social ingredient in the dynamics of
the overall market.

\bigskip
\acknowledgements

We are grateful for constructive comments from Ian Dodd and Anders
Johansen.  This research was supported in part by the ``Models of
Life" Center under the Danish National Research Foundation. R.D.
acknowledges support from CNPq and FAPERJ (Brazil).






\begin{thebibliography}{99}
\bibitem{optihori}
Simonsen I, Jensen MH and  Johansen A, {\it Optimal Investment Horizons}
2002 Eur. Phys. Journ. {\bf 27}, 583-586

\bibitem{gainloss}
Jensen MH, Johansen A and Simonsen I, 
{\it Inverse Statistics in Economics: The gain-loss asymmetry }
2003 Physica A {\bf 234}, 338-343

\bibitem{invfx}
Jensen MH, Johansen A, Petroni F and Simonsen I, 
{\it Inverse Statistics in the Foreign Exchange Market}
2004 Physica A {\bf 340}, 678-684

\bibitem{Book:Bouchaud-2000}
Bouchaud J-P and Potters M, 
{\sl Theory of Financial Risks: From Statistical
Physics to Risk Management}
(Cambridge University Press, Cambridge, 2000).

\bibitem{Book:Mantegna-2000}
Mantegna RN and Stanley HE, 
{\sl An  Introduction to Econophysics: Correlations and Complexity in
Finance} (Cambridge University Press, Cambridge, 2000).

\bibitem{Book:Hull-2000}
Hull J, 
{\sl Options, Futures, and other Derivatives}, 4th ed.
(Prentice-Hall, London, 2000).

\bibitem{Book:Johnson-2003}
Johnson NF, Jefferies P and Hui PM,
{\sl Financial Market Complexity} (Oxford University Press, 2003).

\bibitem{Johansen}
Johansen A, Simonsen I and Jensen MH, 
Optimal Investment Horizons for Stocks and Markets,
preprint, 2005.

\bibitem{farmer}
Farmer JD, 
{\it Physicists Attempt to Scale the Ivory Towers of Finance}
1999 Comp. in Science and Eng. {\bf 1} (6) 26-39

\bibitem{mandelbrot62}
Mandelbrot BB, 
{\it The Variation of Certain Speculative Prices},
1963 Journal of Business, {\bf 36}, 307-332
 
\bibitem{menger}
Menger C, {\sl Principles of Economics} (Libertarian Press, Grove City,
1994).

\end{thebibliography}
\end{document}